\documentclass[nofootinbib,twocolumn,showpacs,superscriptaddress,prl]{revtex4}
\usepackage{latexsym,amsmath,amssymb}
\usepackage{hyperref}

\def\gfive{\hat\g}

\def\sutwo{SU(2)}
\def\real{{\cal R}}
\def\sufour{SU(4)}
\def\sutwol{\sutwo_L}
\def\sutwor{\sutwo_R}

\def\sothreeone{SO(3,1)}

\def\sofour{SO(4)}
\def\sosix{SO(6)}

\def\soten{SO(10)}

\def\sote{SO(3,11)}

\newcommand\be{\begin{equation}}
\newcommand\ee{\end{equation}}
\newcommand\bea{\begin{eqnarray}}
\newcommand\eea{\end{eqnarray}}

\newcommand\one{{\mathbf{1}}}

\newcommand\s{\sigma}

\renewcommand\t{\theta}

\newcommand\g{\gamma}

\newcommand\q{\ensuremath{{\scriptstyle\,\wedge\,}}}
\newcommand\x{\ensuremath{\times}}
\newcommand\ox{\ensuremath{{\small\,\otimes\,}}}
\newcommand\m{\mu}
\newcommand\n{\nu}

\newcommand\RR[1]{\ensuremath{\mathbf{#1}}}
\newcommand\Rb[1]{\ensuremath{\mathbf{\overline{#1}}}}
\newcommand\C{\ensuremath{\mathbb C}}
\newcommand\R{\ensuremath{\mathbb R}}

\makeatletter
\renewcommand\section{\@startsection {section}{1}{\z@}%
                                   {4ex \@plus 1ex \@minus .2ex}%
                                   {2ex \@plus .2ex}%
                                   {\normalfont\small\bfseries\centering}}
\makeatother

\begin{document}
\addtolength{\abovedisplayskip}{-0.4ex}       
\addtolength{\belowdisplayskip}{-0.4ex}       

\title{Chirality in unified theories of gravity}

\author{F. Nesti} 
\affiliation{Universit\`a dell'Aquila \& INFN, LNGS, L'Aquila, Italy}
\author{R. Percacci} 
\affiliation{
SISSA, via Beirut 4, 34014
Trieste, Italy and 
INFN , sezione di Trieste, Italy}


\begin{abstract} 
  \noindent {\bf Abstract.} We show how to obtain a single chiral
  family of an $SO(10)$ GUT, starting from a Majorana-Weyl
  representation of a unifying (``GraviGUT'') group $SO(3,11)$, which
  contains the gravitational Lorentz group $\sothreeone$. An action is
  proposed, which reduces to the correct fermionic GUT action in the
  broken phase. 
\end{abstract}

\maketitle

\section{Introduction}

\noindent 
Low energy chirality poses strong constraints on unified model
building. For example, in Grand Unified Theories (GUT) the fermionic
multiplet must be in a complex representation of the gauge group. At
the same time, chirality precludes the use of orthogonal groups larger
than $\soten$ or exceptional groups larger than $E_6$ \cite{slansky}.
This is relevant for instance when one tries to put all fermionic
families in a single spinor multiplet~\cite{wilczek,SWZ}. The interplay
between Lorentz and internal representations becomes trickier when
gravity is involved.  In the Kaluza--Klein approach to unification, it
is difficult to obtain chiral fermions in four dimensions, even
starting from chiral representations in higher dimensions
\cite{witten}. In string theory chirality of the low energy degrees of
freedom is achieved by suitably choosing the topology of the compact
dimensions, but then unification comes at the cost of introducing
infinitely many new local degrees of freedom. More recently, an
ambitious attempt to unify all known fields into a single
representation of $E_8$ \cite{lisi} stumbled into chirality
issues~\cite{distler}.

Here we discuss the issue of chirality in the context of theories
where the Lorentz group, which is gauged in theories of gravity, is
unified with a GUT group in a larger group $G$. 
By this we mean that the gravitational connection and the gauge fields
of a GUT are components of a connection for the unifying group $G$.
We will call such a
theory a \hbox{``GraviGUT''} (GGUT). Unlike in \cite{lisi}, we do not
insist on putting all fields in a single representation of $G$:
gravitons, gauge fields, fermions and scalars will belong to different
multiplets. The general idea for this kind of unification has been
discussed in \cite{so14, graviweak1, graviweak2}. It is a rather
natural generalization of the GUT program, encompassing also
gravitational interactions. The main difference is that the order
parameter cannot be a scalar but must include a multiplet of one
forms, called the soldering form.\footnote{In some formulations
  inspired by the Plebanski formalism it may be preferable to use a
  two form, dynamically equivalent to the soldering form on
  shell~\cite{smolin}.}  In~\cite{so14} the use of $G=SO(1,13)$ was
proposed, where the soldering form $\theta^i{}_\mu$ with
$i=1,\ldots,14$ and $\mu=1,2,3,4$ is in the fundamental
representation.\footnote{We observe here that the soldering form used
  in the graviweak unifications~\cite{graviweak1, graviweak2} carried
  the tensor product of two vector representations, because the
  fermions were in the vector representation of the
  group. In~\cite{so14} as well as in the present work the fermions
  are in a spinor representation therefore their tensor product
  contains the fundamental as an invariant subspace, and it is
  consistent to restrict oneself to it.}  If the dynamics generates a
VEV for $\theta$ which has rank 4, then one can choose a gauge where
$\theta^i{}_\mu=0$ for $i=5,\ldots,14$. This ``unitary gauge'' breaks
the original gauge group to $SO(10)$, and the breaking scale is
identified with the Planck scale.  The Lorentz and mixed parts of the
connection all become massive at this scale, explaining why we do not
see these degrees of freedom at low energies.

As a preliminary step, in \cite{graviweak1} we discussed mainly the
possibilities for a unification of gravity with the weak interactions.
This 'graviweak' unification is also the basis for a model of
geometrical origin ~\cite{graviweak2} that predicts also the right
strong interactions, but at the price of duplicating the unified
gravitational sector at high energy.  Here we want to include also the
strong interactions in a single unified group.  Probably the most
promising path towards this unification is via the Pati-Salam
model~\cite{patisalam}, based on the group
$\sutwol\times\sutwor\times\sufour$. In view of the fact that this
group is locally isomorphic to $\sofour\x\sosix$, and that the Lorentz
group is also (pseudo)-orthogonal, it seems natural to choose $G$ to
be a pseudo-orthogonal group $SO(p,q)$ with $p+q=14$. In order to
accommodate the Pati-Salam and Lorentz groups, the possibilities are
restricted to $SO(1,13)$, $SO(3,11)$, $SO(5,9)$, $SO(7,7)$. In the
latter two cases the weak and strong gauge fields would belong to
subalgebras with different signature, so that a standard Yang-Mills
action would lead to ghosts. We will restrict our attention to the
remaining two possibilities, which thus contain the full $\soten$ GUT.

It has already been noticed~\cite{so14}, for the case $G=SO(1,13)$,
that the fermion multiplets occurring at low energy lend support to
this unification scenario. In fact the $\RR{64}$, chiral spinor
representation of $SO(1,13)$, breaks under the subgroup $SO(1,3)\times
SO(10)$ into $(\RR{2},\RR{16})\oplus (\Rb{2},\Rb{16})$.  The fact that
the known fermions are spinors of Lorentz and spinors of $SO(10)$
would thus be naturally explained.  Here we will consider in greater
detail the case $G=SO(3,11)$, which admits Majorana-Weyl spinors. In
section II we will show by explicit construction that one such
representation gives rise to a single standard model family, which can
be identified with a chiral $(\RR{2},\RR{16})$.  In so doing we find
the transformations that relates a basis for the Clifford algebra of
$SO(3,11)$ to a basis which is adapted to the subgroup $SO(3,1)\times
SO(10)$. This will allow us, in the section III, to write the kinetic
term for the fermions in an $SO(3,11)$--invariant way, and to see how
it reduces to the familiar one at low energy.  We also show that the
mixed (Lorentz-GUT) gauge fields mediate new high energy processes.
In section IV we conclude with some further comments.

\section{$SO(3,11)$ Spinors and GGUT}

\noindent We start from a set of (128-dimensional, complex) gamma
matrices $\g_i$ for $SO(3,11)$ given explicitly in the Appendix, and
the corresponding chirality operator $\gfive=\Pi_{i=1}^{14}\g_i$ and
algebra generators $\Sigma_{ij}=\frac14[\gamma_i,\gamma_j]$. It is a
property of the Dirac representation that it is equivalent to its
hermitian conjugate, its complex conjugate and its transpose. These
equivalences are realized by three intertwining operators $A$, $B$,
$C$, defined by:
$$
\Sigma_{ij}^\dagger A=-A\,\Sigma_{ij}\,,\quad
\Sigma_{ij}^tC=-C\,\Sigma_{ij}\,,\quad
 B\,\Sigma_{ij}^*=\Sigma_{ij}\,B\,.
$$
The matrices $A$ and $C$ can be used to construct the invariant
hermitian and bilinear forms $\psi_1^\dagger A\psi_2$ and $\psi_1^tC
\psi_2$. The matrix $B$ defines charge conjugation
$\psi^c=(B\circ\star)\psi\equiv B\psi^*$, which for $SO(3,11)$ is an
antilinear
involution because $BB^*=\one$.  One can thus define the left/right
eigenspaces of $\hat\gamma$ by $\hat\gamma\psi_{L/R}=\mp \psi_{L/R}$
and the $+/-$ eigenspaces of charge conjugation by
$(\psi_\pm)^c=\pm\psi_\pm$. An important property of $SO(3,11)$ is
that the matrices $\gfive$ and $B$ commute.  Thus one can define
simultaneous eigenspaces of chirality and charge conjugation, i.e.\
Majorana-Weyl (MW) spinors

It is possible and convenient to choose a basis that is adapted to the MW
representation, in the sense that 
\be
\label{eq:ABC}
A=C=\one_{64}\ox \sigma_1\,,\quad B=\one_{128}\,,
\quad \hat\gamma=-\one_{64}\ox\sigma_3\,.
\ee
In this basis charge conjugation is just complex conjugation, and the
MW spinors are just the real and imaginary parts of chiral spinors: 
$\psi_{L}=\psi_{L+}+i\psi_{L-}$ (and similarly for $R$).
It is then useful to define a map $\real:\C^n\to\R^{2n}$ from complex
$n$-vectors to real $2n$ vectors by $\real
v=(\mathrm{Re}\,v,\mathrm{Im}\,v)^t$, and the inverse map which
associates to the vector $w=(w_1,w_2)^t\in\R^{2n}$ the vector
$\real^{-1}w=w_1+i w_2$.  Using these maps, we can view the MW spaces
either as complex 32-dimensional or real 64-dimensional
representations.
 
We wish to identify a standard model fermion family with a single
MW representation of $SO(3,11)$, for example with the 64 real
degrees of freedom of $\psi_{L+}$. Then, we need to show that,
decomposed as representations of $\sothreeone\times\soten$, these describe
precisely the 32 complex components of a chiral spinor of Lorentz and
chiral spinor of $\soten$, i.e.\ the representation $(\mathbf{2},\mathbf{16})$.

In order to do this, one has to pick half of the components of
$\psi_{L+}$ and use them as real parts of a complex $SO(10)$ spinor,
while the remaining components give the imaginary parts. There is no
natural way of doing this; in fact, any such operation corresponds to
a choice of a complex structure in $\R^{64}$. The simplest choice
would be $\real^{-1}\psi_{L+}$, but one should not expect it to have
simple transformation properties under the subgroup $SO(3,1)\times
SO(10)$. However, there exist a real (64$\times$64) orthogonal
transformation $W_L$ such that $\real^{-1}W_L\psi_{L+}$ do.  To find it, 
we impose that 51 of the $\sote$
generators match those of $\sothreeone\times\soten$ in the respective
(left) Weyl bases:
\be
\hspace*{-1em}
\real^{-1}W_L\Sigma_{L\,ij}^{(3,11)}W_L^t\real=
\left\{\begin{array}{ll}
\Sigma_{L\,mn}^{(3,1)}\ox\one_{16}\ &{\rm for}\ ij=mn\\[1ex]
\one_{2}\ox\Sigma_{L\,ab}^{(10)}\ &{\rm for}\ ij=ab\,.
\end{array}\right.
\hspace*{-1em}
\label{eq:gensplit}
\ee
(We use indices $m,n=1,2,3,4$ and $a,b=5,\ldots14$.) We find that
the matrix $W_L$ is almost completely determined by these equations,
up to a free angle $\alpha$. Note that we do not impose any
requirement on the remaining 40 generators, 
$\Sigma_{L\,m a}^{(3,11)}$, mixing Lorentz and $\soten$ subspaces.

We have thus found the explicit transformation between a single
MW spinor $\psi_{L+}$ of $\sote$ and a Weyl spinor
$\eta_{(\RR2,\RR{16})}$ of $\sothreeone\times\soten$, representing a family in
a $\soten$ GUT:
\be
\eta_{(\RR2,\RR{16})} = \real^{-1}W_L\,\psi_{L+}\,.
\ee
It is useful to observe that the operator $W_L$ is not linear with
respect to the chosen complex structure. Inverting the above relation,
an antilinear part emerges: 
\be
\label{inversion}
\psi_{L+}=W_L^t \real\,\eta_{(2,16)}
=\real (X_W \eta_{(2,16)} + Y_W \eta_{(2,16)}^*)\,.
\ee
where $X_W$ and $Y_W$ are certain complex matrices.
A consequence of this is that not all generators of $SO(3,11)$ can be
realized linearly on the spinors $\eta_{(2,16)}$: by construction the
Lorentz and $\soten$ generators act linearly (they are a representation!)
but the generators that mix Lorentz and $\soten$ turn out to be antilinear. 
For later reference, they can be written as
\be
\label{eq:extra}
{\mathcal R}^{-1}W_L\Sigma_{L\,m a}^{(3,11)}W_L^t {\mathcal R}
=\frac{{\rm e}^{2i\alpha}}{2}
(CA\g_m)_L^{(3,1)} \!\ox\! (C\g_{a})_L^{(10)}\!\circ\!\star\,.
\ee

\medskip

We have obtained a $(\RR2,\RR{16})$ family of fermions
starting from the MW representation $\psi_{L+}$ of $SO(3,11)$. In
order to understand the fate of the other MW representations, we need
two more facts. The first is that, when (\ref{eq:gensplit}) holds for
$\psi_{L+}$, for $\psi_{R+}$ we have
\be
\hspace*{-1em}
\real^{-1}W_R\Sigma_{R\,ij}W_R^t\real=
\left\{\begin{array}{ll}
\Sigma_{L\,mn}\ox\one_{16}\quad&{\rm for}\ ij=mn\\[1ex]
\one_{2}\ox\Sigma_{R\,ab}\quad&{\rm for}\ ij=ab\,,
\end{array}\right.
\hspace*{-1em}
\label{eq:gensplitr}
\ee
with $W_R=W_L$. Therefore $\psi_{R+}$ can be identified with $(\RR2,\Rb{16})$.
Next we introduce the parity operation, in such a way that in the
broken phase it reduces to spatial parity, i.e.\ a matrix that
anticommutes with the three space-like $\gamma$'s. In our MW basis it is:
\be
P_{(3,11)}=i\g_1\g_2\g_3\gfive=\one_{64}\ox \s_2\,,
\ee
where the phase has been chosen so that $P_{(3,11)}^2 = \one$.\linebreak
Since it is imaginary, we see that it exchanges not only the 
Weyl subspaces, but also the Majorana sectors:
$P\psi_{L\pm}=\pm\psi_{R\mp}$. Since spatial parity maps
$(\RR2,\RR{16})$ to $(\Rb2,\RR{16})$, we have the identification of
$(\psi_{L+},\psi_{L-},\psi_{R+},\psi_{R-})$ with
$(\eta_{(\RR2,\RR{16})},
\eta_{(\Rb2,\Rb{16})},\eta_{(\RR2,\Rb{16})},\eta_{(\Rb2,\RR{16})})$.

As a check, the action of $P_{(3,11)}$ on the subspace of the
$SO(3,1)\x SO(10)$ Dirac spinors
$\eta=(\eta_{(\RR2,\RR{16})},\eta_{(\Rb2,\RR{16})})$ is found to be
simply the spacetime parity $\g_4$:
\be
P=\real^{-1} W\, P_{(3,11)}\,W^t\real=\one_{32}\ox\s_2=\one_{16}\ox\g_4\,.
\ee
Thus, in the broken phase parity is inherited by the Lorentz group.

\medskip

Let us pause to discuss the physical meaning of these group theoretic results.
It is instructive to think of them from an $SO(10)$ GUT perspective.
Each family of fermions is a (\RR{2},\RR{16}) complex, chiral representation 
of $SO(3,1)\times SO(10)$, where $SO(3,1)$ is the Lorentz group.
We have shown that the fields in such a representation can be rearranged 
into a real vector and when this is done they are seen to carry not only
a representation of $SO(3,1)\times SO(10)$, but of the larger group $SO(3,11)$.
We have therefore successfully identified a group that can
be used to unify the gravitational and GUT gauge sectors.
The reason why the existence of this group is not evident in the original 
complex form is that the generators
that are not in $SO(3,1)\times SO(10)$ act antilinearly on the fields.
All the generators form nevertheless a perfectly well defined real 
representation, namely the MW \RR{64} of $SO(3,11)$. 

This construction evades the restrictions that chirality of the low energy spectrum
poses on extensions of GUT theories, which were mentioned in the introduction.
First, it is clear that
chiral fermions that are in a real (or pseudoreal) representations of
a GUT group would always lead to a nonchiral theory, therefore
fermions must be in complex representations. Then one has to avoid the
appearance of antifamilies, which would also be in disagreement with
the chirality of the spectrum of the standard model.  It is in fact
not possible to make antifamilies unobservable by giving them a very
large mass ($\gg\,$TeV) because any mechanism giving mass to a chiral
(anti)family at some high energy scale would necessarily break at
least the weak $SU(2)$ symmetry at that scale. Therefore also
antifamiles should have mass near the electroweak scale, where there
are quite strong constraints on their observation.\footnote{Of course,
  if antifamilies were discovered in the future below the TeV scale
  (e.g.~\cite{nath}) this restriction would have to be reviewed.}

We also recall that the problem of antifamilies always arises for
orthogonal GUT groups larger than $SO(10)$. For instance, the MW
representation \RR{128} of the group $SO(16)$ has been used in an
attempt at family unification~\cite{SWZ}. Under the breaking
$SO(16)\to SO(6)\times SO(10)$ it decomposes as $\RR{128} \to
(\RR4,\RR{16}) \oplus (\Rb4,\Rb{16})$, where \RR{4} is the chiral
spinor of $SO(6)$.  The second factor represents four \Rb{16}
multiplets of the same Lorentz chirality of the \RR{16}, i.e.\ four
antifamilies, showing that the theory is nonchiral. A further problem
in this model is that for $SO(16)$ a mass term of the form
$\psi_{L+}^tC_{(3,1)}C_{(16)}\psi_{L+}$ is allowed, because the matrix
$C_{(16)}$ is block-\emph{diagonal} in Dirac space. Thus one needs
additional symmetries to protect the spinors from a large (Planck or
GUT-size) mass term.

For the MW spinors of the $SO(3,11)$ GGUT suggested here, these
problems are both absent. First, because Lorentz is included in the
unification, the real representation of the GGUT group is actually a
single complex representation of the GUT group. Second, any bare mass
term is forbidden. This can be seen directly as a consequence of
chirality of $SO(3,11)$: because the matrix $C_{(3,11)}$ is
block-\emph{anti}diagonal in Dirac space, then
$\psi_{L+}^tC_{(3,11)}\psi_{L+}^{\phantom{t}}=0$.\footnote{Also
  vanishing because $C_{(3,11)}$ is symmetric while $\psi$ are
  anticommuting.}

Thus, we have shown that GGUTs can be chiral by construction, in spite
of adopting real representations and orthogonal groups larger than
$SO(10)$. In particular, chirality of the GGUT representation is
maintained at low energy. By using $SO(3,11)$ and its Majorana-Weyl
representation one can achieve a single standard model family,
while by using the chiral representation of $SO(1,13)$ one could generate
\emph{two} standard model families. It is also clear that if we
started from a nonchiral (Dirac) representation of $SO(3,11)$ we would
have ended with two families and two antifamilies.

\section{Dynamics}

\noindent 
Constructing an action for a GGUTs poses new challenges that go beyond
those familiar in GUTs. One would like to have an action which is well
defined both in the symmetric and broken phase of the theory.  But in
these theories the symmetric phase is topological (the metric
$\theta^i{}_\mu\theta^j{}_\nu\eta_{ij}$ vanishes classically) so one
cannot use the standard type of actions.  In \cite{so14} a sort of
mean field dynamics was proposed, generating the VEV of $\theta$
selfconsistently.  Another approach is to use techniques which have
been studied in the context of topological theories.  We concentrate
here only on the action for the fermions.

We begin by defining the $SO(3,11)$ covariant derivative
acting on MW spinors
\be
D_\mu\psi_{L+}
=\left(\partial_\mu+\frac{1}{2}A_\mu^{ij}\Sigma_{L+\,ij}^{(3,11)}\right)\psi_{L+}\,.
\ee
Note that $\Sigma_{L\pm\,ij}^{(3,11)}=\Sigma_{L\,ij}^{(3,11)}$ are
real.  Then we define the covariant differential $D$, mapping spinors
to spinor-valued one forms: $D\psi_{L+}=D_\mu\psi_{L+}dx^\mu$. The
quadratic form
\be
\label{eq:quad}
\psi_{L+}^\dagger (A\g^i)_LD\psi_{L+}
\ee
is manifestly a vector under $\sote$ and a one form under
diffeomorphisms.\footnote{The product $A\gamma^i$ is block diagonal in
  Dirac space, because both $A$ and $\g_i$ are block
  \emph{anti}-diagonal.}  Then, to construct a $\sote$-invariant
action, we introduce an auxiliary field $\phi_{ijk\ell}$ transforming
as a totally antisymmetric tensor. The action is
\be
\label{eq:action}
\mathcal{S}=\int\psi_{L+}^\dagger (A\g^i)_LD\psi_{L+}\,\q\theta^j\q\theta^k\q\theta^\ell \,
\phi_{ijk\ell} \,.
\ee

The breaking of the $SO(3,11)$ group to the Lorentz and $SO(10)$
subgroups is induced by the VEV of two fields: the soldering one-form
$\t_\m^i$ and the four-index antisymmetric field
$\phi_{ijk\ell}$.\footnote{The field $\phi_{ijkl}$ also appears in
  Plebanski reformulations of General Relativity, where the vierbein
  field is traded for a two form field. If the (Lorentz) gauge group
  is extended, $\phi$ serves, as in the present context, to achieve
  the symmetry breaking~\cite{smolin}.}  We assume that the VEV of
$\phi_{ijk\ell}$ is $\epsilon_{mnrs}$, the standard four-index
antisymmetric symbol, in the Lorentz subspace, and zero otherwise. The
VEV of the soldering form on the other hand has maximal rank (four)
and is also nonvanishing only in the Lorentz subspace, $m=1,2,3,4$:
\be
\left\{\begin{array}{l}
\phi_{mnrs}=\epsilon_{mnrs}\\
\phi_{ijk\ell}=0 \quad\text{otherwise}
\end{array}
\right.
\quad 
\left\{\begin{array}{l}
\t_\m^m=M e^m{}_\mu\\
\t_\m^a=0 \quad\text{otherwise}
\end{array}
\right.
\label{eq:VEVs}
\ee
where $e^m{}_\mu$ is a vierbein, corresponding to some solution of the
gravitational field equations which we need not specify in this
discussion (below we will choose $e^m{}_\mu=\delta^m_\mu$) and $M$ can
be identified with the Planck mass.  Clearly the breaking pattern just
described is the one that leads to a theory which is Lorentz invariant
(at each point) but other choices may be possible (see comments below).


Using (\ref{inversion}) and omitting the subscript $(\RR2,\RR{16})$
from the spinors, the kinetic quadratic form (\ref{eq:quad}) becomes
\be
\eta^\dagger \,\real^{-1}W_L (A\gamma^i)_L D W_L^t\real\,\eta\,.
\ee
In the broken phase, treating separately the cases $i=m=1,2,3,4$ 
and $i=a=5,\ldots14$, we find:
\bea
\label{eq:gammasplit}
\!\!\!\!
&&\real^{-1}W_L (A\gamma^m)_L W_L^t\real=
i(A\gamma^m)^{(3,1)}_L\!\ox\!\one_{16}\,\\
\!\!\!\!
&&\real^{-1}W_L (A\gamma^a)_L W_L^t\real=
i {\rm e}^{2i\alpha} 
C_L^{(3,1)}\!\ox\! (C\gamma^a)^{(10)}_L\!\circ\!\star\,.\ \ \ \ 
\eea
Therefore, using (\ref{eq:gammasplit}) and the fact that for Lorentz 
$(A\gamma^m)^{(3,1)}_L=\s^m$, together with (\ref{eq:gensplit})
for the connection terms in the covariant derivative, the action with
a flat background vierbein reduces to the standard one for a $SO(10)$
family in flat space:
\be
\label{eq:action}
\int d^4x\,\eta^\dagger\sigma^\mu \nabla_\mu\eta \,,
\ee
where now
$\nabla_\mu=D^{(10)}_\mu=\partial_\mu+\frac{1}{2}A_{\m\,(10)}^{ab}\Sigma^{(10)}_{ab}$
is the $SO(10)$ covariant derivative. Note that this action contains
the standard kinetic term of the fermions, and the interaction with
the $SO(10)$ gauge fields, which at this stage can still be assumed to
be massless.

Had we chosen a nonflat gravitational background, the action would
contain the invariant volume factor $|e|$ and the covariant derivative
would also contain a nontrivial Lorentz part:
$\nabla_\mu=D^{(10)}_\mu+\frac{1}{2}A_{\mu\,(3,1)}^{mn}\Sigma^{(3,1)}_{mn}$.
As discussed in \cite{so14}, the Lorentz connection
$A_{\mu\,(3,1)}^{mn}$ in the covariant derivative can be assumed to be
the Levi-Civita connection derived from the vierbein. Its fluctuations
around this VEV are also present but have a mass of the order of the
Planck mass and are negligible at low energies.

The remaining $A_\m^{ma}$ components of the $SO(3,11)$ connection,
that mix Lorentz and $SO(10)$, also have Planck mass. These gauge
fields, carrying a Lorentz and a $\soten$ vector index, can be
decomposed in Lorentz representations by lowering the $m$ index with a
vierbein, leading to the two fields $A_{(\m\n)}^a$, $A_{[\m\n]}^a$.
They contain thus a symmetric and an antisymmetric field, both in the
representation $\RR{10}$ of $SO(10)$, that interact with fermions via
the following vertex:
\bea
\label{eq:vertex}
&&\!\!\!\!{\rm e}^{2i\alpha}A_\m^{ma} \, \eta^t [(C\g^\m\g_m)_L^{(3,1)} \ox (C\g_{a})_L^{(10)}]\eta=\\[-.2ex]
&&= {\rm e}^{2i\alpha} \, \eta^t [ C^{(3,1)} (A_{(\m\n)}^{a} g^{\m\n}+A_{[\m\n]}^{a}\s^{\m\n})\ox (C\g_{a})_L^{(10)}]\eta\,.
\nonumber
\eea
%
The first of the two vertices is equivalent to the one generated by
the standard scalar Higgs field \RR{10} of $SO(10)$, while the second
is a new vertex that involves the spin. The resulting four fermion
interactions may lead to new gravitational contributions to rare
processes.

We observe that even though these new interactions originate from the
generators mixing $SO(10)$ and Lorentz indices, if the breaking works
as above, global Lorentz symmetry is not broken by
these interactions, because the original lagrangian has local Lorentz
symmetry as a (subgroup of the) gauge symmetry, and the background
VEVs (\ref{eq:VEVs}) preserve the global spacetime remnant of this
gauge symmetry (see~\cite{graviweak1,mixing} for a detailed discussion).

\section{Summary and Outlook}

\noindent
A GraviGUT is a very natural extension of a GUT, encompassing also
gravitational interactions.  Given that the (pseudo)orthogonal group
plays a fundamental role in the theory of gravity, it is especially
attractive to consider GGUTs that are (pseudo)-orthogonal extensions 
of an $SO(10)$ GUT.
The minimal theory of this type can be based on $SO(1,13)$ or $SO(3,11)$.
We have shown that the latter choice is slightly more natural
from the point of view of the fermionic content, because it can
accommodate three families, whereas $SO(1,13)$ leads to an even number
of families.
The field content of the simplest GGUT would thus be an $SO(3,11)$ Yang-Mills field,
three Majorana-Weyl fermions plus whatever is needed to break the original 
symmetry to what we see at low energy.
The first step of the symmetry breaking chain is essentially unique:
$SO(3,11)\to SO(3,1)\times SO(10)$.
This is achieved by postulating a nontrivial VEV for a suitable order parameter.
The distinctive feature of this first step is that the order parameter
is not a scalar but rather a one form with values in the vector 
representation of the gauge group, $\theta^i_\mu$.
This so called soldering form provides the necessary connection
between spacetime and internal transformations, and its first four components
$\theta^m_\mu$ carry the gravitational degrees of freedom in the broken phase.

At this stage it is less clear what degrees of freedom are needed to
describe the further breaking of $SO(10)$ to the standard model group
$SU(3)\times SU(2)\times U(1)$,
and the final breaking of the latter to the electromagnetic $U(1)$.
This will have to be investigated in the future.
In principle requiring the GGUT representations to decompose 
into well-behaved states at low energy,
together with the restrictive choice of a GGUT group, 
should pose constraints also on the GUT sector.
At the same time we observe that the breaking of the GUT group 
is anyway an open issue (see e.g.~\cite{bertolini} for a recent thorough 
reanalysis of non-SUSY $SO(10)$), and that even in the context of the
Standard Model the origin of the electroweak symmetry breaking is still
partly shrouded in mystery. So it should not come as too much of a
surprise if this sector of the GGUT is also less understood.

In the present paper we have discussed in detail the kinematics (sect. II) 
and dynamics (sec. III) of the fermionic sector.
In particular, in section II, we have shown explicitly the equivalence
beween the MW representation $\RR{64}$ of $SO(3,11)$ and the
$(\RR2,\RR{16})$ chiral, complex spinor representation of Lorentz
and $SO(10)$, representing a family of Standard Model fermions.
This identification evades the problems that chirality of the Standard
Model spectrum poses to unified theories, and thus $SO(3,11)$ can be
safely adopted as a basis for a unified theory.
A further consequence of this construction is that
$SO(3,11)$ is also the largest (pseudo-orthogonal) group allowing a chiral low
energy spectrum, and thus attempts to achieve family unification 
by further enlargement of the group are
not possible in this approach without introducing mirror families.
In section III we have then constructed a diffeomorphism- and
$SO(3,11)$-invariant action for fermions and shown how, under a
suitable symmetry breaking realized by means of the soldering form and
an additional antisymmetric tensor field, this reduces to the correct
$SO(10)$-invariant action coupled to gravity at low energy.  
Various hurdles will have to be overcome in the development of GGUTs,
but we have shown here that the construction of a realistic fermionic
sector is not an obstacle.
We can thus claim that, at least on this count,
the setup described here represents the first realistic
framework that unifies gravity with the other known interactions.
In the rest of this section we discuss a few of the open issues.

The bosonic part of the action, including the gauge and Higgs terms, 
is probably the most important omission.
In a less ambitious form of unification, it has been discussed
in~\cite{graviweak1,graviweak2}, see also~\cite{smolin,ss,krasnov}
and, for a completely different approach, \cite{so14}. 
In this connection, an issue that is sometimes raised is the presence of ghosts:
given that the gauge group is noncompact, one expects that some
components of the connection will have wrong sign kinetic terms.
Surely, one wants to avoid ghosts in the low-energy GUT gauge sector:
this problem was already mentioned in the introduction, and we used it to
select some group rather than others. The GGUT groups we discarded
would have led to ghosts with a mass of the order of the GUT or lower,
while the groups we selected would seem naively to have ghosts with Planck mass. 
This is what happens also in generic gravitational theories with propagating torsion, 
independent of unification~\cite{vann}. Over time, there have been various proposals
to circumvent this problem~\cite{LeeWick,SalamStrathdee}.  
Here we may add that since the ghosts would occur near or beyond the 
transition 
to a different, topological phase, the standard tree level analysis is 
certainly not conclusive.

The detailed phenomenology of a GGUT will depend upon the details
of the symmetry breaking chain. 
As in ordinary GUTs, the most characteristic signal will come
from new interactions mediated by the components of the gauge field on
the broken generators of the GGUT group, in the present case the heavy
gauge fields mixing Lorentz and $SO(10)$ indices.
Their effect is similar to that of a $SO(10)$ higgs field in the representation $\RR{10}$.  
We have shown that the corresponding generators are antilinear, 
and these processes will violate fermion number by two units.
One can expect that interactions similar to proton decay and
neutron-antineutron oscillations would be present, but with new spin
structure. These interactions would be suppressed by the large mass of $A_\mu^{ma}$,
so only extremely rare processes would have a chance of being observable.

The symmetry breaking VEVs that we proposed conserve the Lorentz
symmetry, but it is conceivable that, with different VEVs of $\t$ and
$\phi$, Lorentz symmetry could be broken (even locally) as it happens
in theories with more tensor condensates. This may lead to Lorentz
violation in proton decay (as first discussed in~\cite{LBproton}), a
striking possibility since proton decay experiments have assumed so
far strict Lorentz invariance, possibly missing already occurring
events. On the other hand, the coupling of both $\t$ and $\phi$ to
fermions may introduce such a Lorentz-symmetry breaking also in the
matter sector.

Another major issue that we did not mention so far is that a proper
understanding of the GGUT breaking mechanism will require a theory of
quantum gravity.  It is clear that at sufficiently low energy the
Planck mass fields decouple and that the remaining ones can be
described by an effective field theory.  We are assuming that adding
the Planck mass fields one can somehow obtain a well defined quantum
theory.  Asymptotic safety could be of help here, see \cite{as} and
references therein.  We note finally that if unification works as
described here, the mystery of the origin of flavors appears to be
even deeper than the issues posed by quantum gravity.


\section{Acknowledgements}

Work partially supported by the ASI contract I/016/07/0 ``COFIS'' and
by the European FP6 Network ``UniverseNet'' MRTN-CT-2006-035863. We
would like to thank Z. Berezhiani, G. Senjanovic for discussions.

\section{Appendix}

A Weyl basis for Euclidean $SO(n)$ gamma matrices can be constructed
recursively for even $n$ starting from $n=2$ with $\g_{2,1}=\sigma_1$,
$\g_{2,2}=\sigma_2$, using the rules
\bea
\g_{n,i}&=&\g_{n-2,i}\,\gfive_{n-2}\ox (-i\s_2) \qquad \text{for}\ i<n-1\nonumber\\[.7ex]
\g_{n,n-1}&=&\one_{d(n-2)}\ox\sigma_1\,,\nonumber\\[.7ex]
\g_{n,n}&=&\gfive_{n-2}\ox \s_2\,,
\eea
where $d(n)=2^n$ is the dimension of the representation and
$\gfive_n=(-i)^{n/2}\Pi_{i=1}^n\g_{n,i}$ is the chirality matrix.  As
one checks, it has the right form
${\hat\gamma}_n=\one_{d(n-2)}\ox\s_3$.  The generators of the algebra
are
\be
\Sigma_{n,ij}=\frac{1}{4}[\g_{n,i},\g_{n,j}]\,,
\ee
and are antihermitian and block diagonal.

\medskip

In signature $(3,11)$ ($3$ negative, $11$ positive eigenvalues)
the gamma matrices are given by
\be
\g_k=\left\{\begin{array}{ll}
i \g_{14,k}\qquad & {\rm for}\ 1\leq k\leq 3\,,\\[0.7ex]
\g_{14,k}\qquad& {\rm for}\ 3<k\leq 14
\end{array}\right.
\ee
and the definition of $\gfive$ has an additional factor $i^3$
so that $\gfive=\Pi_{i=1}^{14}\g_i$.  The conjugation operations are
\be
A=\g_1\g_3\,,\quad 
B=\g_1\g_3 \g_4\g_6\g_{8}\g_{10}\g_{12}\g_{14}\,,
\ee
and $C=B A^*$.

\pagebreak[3]

The explicit gamma matrices are: 
{\arraycolsep=1.2pt
\be
\begin{array}{ccrcccccccccccccccccc}
\g_1&=& i&\s_2&\ox&\s_1&\ox&\s_1&\ox&\s_1&\ox&\s_1&\ox&\s_1&\ox& \s_2\\
\g_2&=&-i&\s_1&\ox&\s_1&\ox&\s_1&\ox&\s_1&\ox&\s_1&\ox&\s_1&\ox& \s_2\\
\g_3&=&-i& \one&\ox&\s_2&\ox&\s_1&\ox&\s_1&\ox&\s_1&\ox&\s_1&\ox& \s_2\\
\g_4&=&-&\s_3&\ox&\s_1&\ox&\s_1&\ox&\s_1&\ox&\s_1&\ox&\s_1&\ox& \s_2\\
\g_5&=&-& \one&\ox&\one&\ox&\s_2&\ox&\s_1&\ox&\s_1&\ox&\s_1&\ox& \s_2\\
\g_6&=&  & \one&\ox&\s_3&\ox&\s_1&\ox&\s_1&\ox&\s_1&\ox&\s_1&\ox& \s_2\\
\g_7&=&-& \one&\ox&\one&\ox&\one&\ox&\s_2&\ox&\s_1&\ox&\s_1&\ox& \s_2\\
\g_8&=& & \one&\ox&\one&\ox&\s_3&\ox&\s_1&\ox&\s_2&\ox&\s_1&\ox& \s_2\\
\g_9&=&-& \one&\ox&\one&\ox&\one&\ox&\one&\ox&\s_2&\ox&\s_1&\ox& \s_2\\
\g_{10}&=& & \one&\ox&\one&\ox&\one&\ox&\s_3&\ox&\s_1&\ox&\s_1&\ox& \s_2\\
\g_{11}&=&-&\one&\ox&\one&\ox&\one&\ox&\one&\ox&  \one  &\ox&  \s_2  &\ox& \s_2\\
\g_{12}&=& &\one&\ox&\one&\ox&\one&\ox&\one&\ox&  \s_3  &\ox&  \s_1  &\ox& \s_2\\
\g_{13}&=&  &\one&\ox&\one&\ox&\one&\ox&\one&\ox&  \one  &\ox&  \one  &\ox& \s_1\\
\g_{14}&=&  &\one&\ox&\one&\ox&\one&\ox&\one&\ox&  \one  &\ox&  \s_3  &\ox& \s_2
\end{array}
\ee
so that one finds:
\bea
A&=& -\s_3\ox\s_2\ox\s_1\ox\s_1\ox\s_1\ox\s_1\ox\s_2\\
B&=& +\s_1\ox\s_1\ox\s_3\ox\s_2\ox\s_3\ox\s_2\ox\one\\
C&=& -\s_2\ox\s_3\ox\s_2\ox\s_3\ox\s_2\ox\s_3\ox\s_2
\eea
that are real, symmetric (hermitian) and orthogonal.

\medskip

The Weyl basis described is not unique, and any similarity $\g'= S\g
S^{-1}$ with $[S,\gfive]=0$, preserving the algebra and the Weyl form,
trasforms the conjugations as:
\be
A'= S A S^\dagger\,,\ B'= S B S^{*-1}\,,\ C'= SCS^T\,.
\ee
This freedom has been exploited in the text to adapt the basis to the
MW representation and reach the form~(\ref{eq:ABC}). In particular we used
\be
S^{-1}=S_BS_AS_M\,,
\ee
with 
\bea
S_B&=&[\s_1\ox\s_1\ox\s_1\ox\s_1\ox\s_1\ox (\s_1-i\s_2-\s_3+\one)-\nonumber\\
     &&   \one\ox\one\ox\s_2\ox\s_3\ox\s_2\ox(\s_1+i\s_2-\s_3-\one)]\ox\one\,,\nonumber\\
S_A&=&\one_{64}\ox(\one-\s_3)+F\ox(1+\s_3)\,,\nonumber\\
S_M&=&\one_{32}\ox(\one_4+i\s_3\ox\one)\,.
\eea
where $F=i\s_2\ox\s_3\ox\s_2\ox\s_3\ox\s_2$. ($S_B$ and $S_A$ bring $B$
and $A$ to diagonal form, and $S_M$ brings $B$ to the identity.)  }

A twist was also adopted to reach a standard basis in signature $(3,1)$.

\raggedright

\newpage

\end{document}